\documentstyle[12pt,aasms4]{article}

\lefthead{Kaper et al.}
\righthead{Bow shock around Vela~X--1}

\newcommand{\kms}{km~s$^{-1}$}
\newcommand{\vela}{Vela~X--1}
\newcommand{\ha}{H$\alpha$}

\begin{document}

\title{Discovery of a bow shock around Vela~X--1}

\author{L.~Kaper\altaffilmark{1}, J.Th.~van~Loon\altaffilmark{1}, 
T.~Augusteijn\altaffilmark{2}, P.~Goudfrooij\altaffilmark{1}, 
F.~Patat\altaffilmark{1}, L.B.F.M.~Waters\altaffilmark{3}, 
A.A.~Zijlstra\altaffilmark{1}}
\altaffiltext{1}{European Southern Observatory, 
Karl-Schwarzschild-Strasse 2, D--85748 Garching, Germany}
\altaffiltext{2}{European Southern Observatory, 
Casilla 19001, Santiago 19, Chile}
\altaffiltext{3}{Astronomical Institute ``Anton Pannekoek'',
University of Amsterdam, Kruislaan 403, 1098 SJ Amsterdam, The Netherlands}

\begin{abstract}
We report the discovery of a symmetric bow shock around the
well-known high-mass X-ray binary (HMXB) Vela~X--1. Wind bow shocks
are a ubiquitous phenomenon around OB-runaway stars, but now
such a structure is found around a HMXB. The presence of a bow shock
indicates that the system has a high (supersonic) velocity with
respect to the interstellar medium. From the symmetry of the bow
shock, the direction of motion and, moreover, the origin and age of
the system can be derived.  Our observation supports Blaauw's scenario
for the formation of an OB-runaway star by the supernova explosion of 
the binary companion.
\end{abstract}


\section{Introduction}

The high-mass X-ray binary (HMXB) HD77581 (Vela X-1) consists
of a B0.5~Ib supergiant (Vidal et al. \cite{VW73}, Jones \& Liller
\cite{JL73}) and an X-ray pulsar companion ($P_{\mbox{pulse}}$=283s,
McClintock et al. \cite{MR76}) in an 8.9 days orbit.  Pulse-timing
analysis of the X-ray pulsar has resulted in the accurate
determination of its orbital parameters (Nagase \cite{Na89}). The
length of the X-ray eclipses (Watson \& Griffiths \cite{WG77})
indicates that the inclination of the orbital plane is close to
90$^{\circ}$. Also the radial-velocity curve of the supergiant
companion is measured (Van Paradijs et al. \cite{PZ77}, Van Kerkwijk
et al. \cite{KP95a}), so that the masses of both binary components
($23.5^{+2.2}_{-1.5}$ and $1.88^{+0.69}_{-0.47}$~M$_{\odot}$,
3$\sigma$ limits) can be derived. It turns out that \vela\ is the most
massive neutron star known (Van Kerkwijk et al. \cite{KP95b}),
although the 3$\sigma$ error bar still includes the canonical neutron
star mass of 1.4~M$_{\odot}$.

Evolutionary scenarios for massive binaries (see e.g.\ Van den Heuvel
\cite{He93} for a review) predict that the HMXB \vela\ originates from
a 25+22.5 M$_{\odot}$ binary (cf.\ Van Rensbergen et al. \cite{RV96}).
When the initially most massive star (the primary) becomes a
supergiant and starts filling its Roche lobe, a phase of mass transfer
occurs, rejuvenating the secondary that eventually becomes the most
massive star in the system. The primary evolves into a helium star
that leaves a neutron star (or a black hole) after a supernova
explosion. When less than half of the total mass of the system is
lost, the system (now containing an OB star and a compact companion)
remains bound (Blaauw \cite{Bl61}) and gets a kick velocity. The
binary system then leaves the location where it was born.  As soon as
the secondary OB star evolves into a supergiant, accretion of its
dense stellar wind onto the compact star will power a strong X-ray
source. Thus, according to this model of massive binary evolution, all
HMXBs should be runaways.  

\subsection{Blaauw's scenario}
  
Zwicky (\cite{Zw57}) and Blaauw (\cite{Bl61}) were the first to
realize that a supernova explosion of one of the binary components can
result in a high space velocity of the companion star. This mechanism
for the formation of OB-runaway stars is presently known as Blaauw's
scenario.  In the original scenario a phase of mass transfer was not
included, so that the massive binary had a higher probability to
disrupt after the supernova explosion of the primary. But the modern
version of this scenario predicts that the resulting runaway star has
a high probability to remain bound to its compact companion (Van den
Heuvel \cite{He93}, Hills \cite{Hi83}). Although many people expect
that a significant fraction of OB~runaways has a compact companion,
searches for compact stars around OB~runaways have up to now not been
successful (e.g.\ Philp et al. \cite{PE96}).

As a result of a phase of mass transfer in a compact binary, one might
expect that the abundances of nuclear processed elements are enhanced
in the atmosphere of an OB-runaway star. Also, the angular momentum
associated with the accreted material will increase the rotation rate
of the OB runaway.  For a small sample of bright OB runaways, Blaauw
(\cite{Bl93}) shows that these stars indeed exhibit a tendency towards
higher helium abundance and higher projected rotational velocity. He
further argues that some OB runaways seem to be younger than the age
of the OB association they originate from (these stars are called blue
stragglers). These observations provide, however, only circumstantial
evidence for the supernova origin of (some) OB runaways.

\subsection{Cluster ejection mechanism}

An alternative explanation for the existence of OB-runaway stars is
the cluster ejection model (Poveda et al. \cite{PR67}): dynamical
interaction in a compact cluster of stars results in the ejection of
one or more of the members. From their extensive radial velocity
survey of bright OB-runaway stars, Gies \& Bolton (\cite{GB86})
concluded that the cluster ejection model should be favoured. Their
conclusion is mainly based on arguments refuting a supernova origin
for the observed OB-runaway stars. Apart from the lack of
observational evidence for the presence of compact companions around
OB runaways, the existence of 2 runaway double-lined spectroscopic
binaries cannot be explained with the supernova model. Also the
kinematical age of OB-runaway stars (i.e.\ the time needed to reach
its present position with respect to the ``parent'' OB association) is
often close to the age of the OB association itself, which is in
support of the cluster ejection model. On the basis of the observed
radial velocities they argued that the population of HMXBs and
OB-runaways appears to be different. Van Oijen (\cite{VO89}), however,
found strong indications that HMXBs are high velocity objects.

In the following we will provide compelling observational evidence
that at least one HMXB is an OB~runaway system, that obtained its high
space velocity through the supernova explosion of the compact star's
progenitor. 

\section{Wind bow shock around Vela~X--1}

Kinematic studies of OB stars are hampered by the large distances at
which these stars are usually found, making it very difficult to
measure proper motions accurately (although this situation will be
significantly improved after release of the Hipparcos data). But
it turns out that many OB stars with high space velocity create an
unmistakable sign in the surrounding space.  When an OB~star moves
supersonically through the interstellar medium (ISM), the interaction
of its stellar wind with the ISM gives rise to a bow shock. Van Buren
\& McCray (\cite{BM88}) examined the Infra Red Astronomical Satellite
(IRAS) all-sky survey at the location of several OB-runaway stars and
found extended arc-like structures associated with many of them. The
infrared emission results from interstellar dust that is swept up by
the bow shock and heated by the radiation field of the OB~star. In a
subsequent study (Van Buren et al. \cite{BN95}), wind bow shocks were
detected around one-third of a sample of 188 candidate OB-runaway
stars. Thus, the detection of a wind bow shock can be considered as an
observational confirmation of the runaway status of an OB~star.

To this aim, we searched for the presence of a wind bow shock around
the HMXB \vela. This is a good candidate for an OB~runaway system.
Its runaway nature was proposed in a recent paper by
Van~Rensbergen et al. (\cite{RV96}), based on the observed annual
proper motion of HD77581 listed in the Hipparcos Input Catalogue
(Turon et al. \cite{Tu92}, see Table~1).  The direction of the annual
proper motion suggests that HD77581 might originate from the
OB~association Vel~OB1, which is located at a distance of 1820 parsec
(Humphreys \cite{Hu78}) from the Sun. At that distance, HD77581
(\vela) would have a space velocity of about 90~\kms\ and would have
left Vel~OB1 $2 \pm 1$ million years ago. They further showed that
this kinematical age is consistent with the predicted time of
supernova explosion of Vela~X-1's progenitor.

Figure~1 (Plate~X) shows a narrow-band \ha\ image of a $10 \times 10$
arcminute-squared field centered on \vela, obtained with the 1.54m
Danish telescope at the European Southern Observatory (ESO) on
February 14, 1996. The bright star HD77581 (V=6.9) has been removed
(including its reflections) from the \ha\ frame (exposure time 20
minutes) by subtracting a properly scaled R-band image of the
same field. The uncalibrated \ha\ image clearly reveals the presence
of a bow shock, the apex being at $0.9 \pm 0.1$ arcminute to the north
of HD77581. At the distance of HD77581 (see below), this corresponds
to a projected distance of $0.48 \pm 0.05$ parsec. With help of the
flux-calibrated long-slit spectrum (see Fig.~2) and knowledge of the
position and dimensions of the slit, we estimate (within a
factor of two) the image peak H$\alpha$ intensity to be $\sim
10^{-15}$ erg~s$^{-1}$~cm$^{-2}$~arcsec$^{-2}$. The western (right)
part of the bow shock consists of two separate filaments. Obviously,
one filament is the continuation of the bow shock to the
west. Filamentary structure has been observed in other wind bow shocks
as well and might be related to their instability (Dgani et
al. \cite{DV96}). The IRAS small-scale structure catalogue (Helou \&
Walker \cite{HW88}) lists an extended infrared source at the position
of \vela. The extended source can be identified as an H~{\sc ii}
region with a Str\"{o}mgren radius of $\sim 0.2$ degrees ($\sim 7$
parsec). We produced a high-resolution IRAS map (not shown here) of
this area on the sky and found that inside the H~{\sc ii} region the
bow shock is well resolved at infrared wavelengths too, most
pronounced in the 60$\mu$m wavelength band.

\section{Bow shock structure}

The structure of a wind bow shock is determined by the balance between
the ram pressures of the wind and the ambient medium (Baranov et al.
\cite{BK71}). One can show that:
\[ 4 \pi \rho_{a} v_{\star}^{2} R_{0}^{2} = \mbox{\.{M}}_{w} v_{w} \, . \]
Using the observed value for the distance $R_{0}$ between the OB~star
and the stagnation point (apex), the mass-loss rate $\mbox{\.{M}}_{w}
= 10^{-6}$ M$_{\odot}$~yr$^{-1}$ and terminal velocity $v_{w} = 1105$
\kms\ of the stellar wind of HD77581 (Kaper et al. \cite{KH93}), and
90 \kms\ for the space velocity $v_{\star}$ of the system, we derive
for the density of the ambient medium: $\rho_{a} \approx 2.6
\times 10^{-24}$~g~cm$^{-3}$ (number density $\sim 1$ cm$^{-3}$). 

A rough estimate of the local ISM density can be obtained from the
size of the H~{\sc ii} region and also from its observed infrared
flux. Following Osterbrock (\cite{Os89}) and assuming that the H~{\sc
ii} region is a pure hydrogen nebula, optically thick in the Lyman
series, a Str\"{o}mgren radius of 7 parsec and a black-body input
spectrum ($T_{\rm eff}$=20.000~K) results in a neutral hydrogen number
density $n_{\rm H}$ of 6~cm$^{-3}$. The total IR flux produced
by dust inside the H~{\sc ii} region is a measure for the number of
Ly$\alpha$ photons produced by recombinations. The flux observed
by IRAS (integrated over the H~{\sc ii} region and corrected for the
IRAS band widths) is $3 \times 10^{-8}$ erg cm$^{-2}$ s$^{-1}$. A
black-body fit to the flux values shows that the dust temperature is
78~K and that the total dust emission is $5 \times 10^{-8}$ erg
cm$^{-2}$ s$^{-1}$, about 1.7 times the flux detected by IRAS. The
energy per second per unit volume produced by recombination to the
ground level is $3.3 \times 10^{-24} \times n_{\rm H}^{2}$ erg
cm$^{-3}$ s$^{-1}$. Comparison with the measured IR flux gives $n_{\rm
H} = 16$~cm$^{-2}$. If the emission from the bow shock is
excluded, this yields $n_{\rm H} = 11$~cm$^{-2}$. Given the
uncertainties, the estimate of the local ISM density with help of the
bow shock and stellar wind parameters is probably the most accurate.

We obtained a long-slit spectrum of the bow shock with the ESO 1.52m
telescope on the night of May 28, 1995 (Figure~2). We show a
flux-calibrated spectrum for two regions covered by the slit; the
orientation and position of the slit is indicated in the sub-panel.
The spectra include the \ha\ and H$\beta$ lines, and forbidden
emission lines like \mbox{[O~\sc{iii}]} $\lambda\lambda$4959,5007~\AA\
and \mbox{[N~\sc{ii}]} $\lambda$6583~\AA. Region A corresponds to the
secondary filament which appears to produce the maximum H$\alpha$
intensity. The H$\alpha$/H$\beta$ ratio in region A is 6.5; if this
large ratio is due to extinction, E(B-V) would be equal to 0.72, which
is consistent with the reddening of HD77581 (Sadakane et
al. \cite{Sa85}). In region B, which covers a part of the wind bow
shock, the Balmer lines are much weaker with respect to the forbidden
\mbox{[O~\sc{iii}]} lines (\mbox{[O~\sc{iii}]} $\lambda$5007/H$\beta$
$\geq 4.1$) than in region A ($\approx 1.5$). Compared to other usual
shock-excited emission lines, the \mbox{[O~\sc{iii}]} lines originate
close to the shockfront where the ionization conditions depend
critically on shock conditions (Dopita \cite{Do77}, Hollis et
al. \cite{HO92}). Therefore, the spectra are in accordance with both
regions being part of the wind bow shock. The filament covered by
region A might be experiencing a somewhat weaker shock, yielding a
much larger emission measure in H$\alpha$.

\section{Discussion and conclusions}

The present position of HD77581 and the OB~association Vel~OB1 is
shown in Figure~3. The separation between the two is about 7
degrees. The direction of the proper motion of HD77581, (a) as listed
in the HIC and (b) as measured by Sahu (\cite{Sa92}), is
indicated. The dotted line represents the proposed path of HD77581
based on the symmetry-axis of the bow shock, which is in a direction
in between the proper motion vectors $\vec{a}$ and $\vec{b}$. The bow
shock reflects the relative motion between the star and the
surrounding medium, while the proper motion measures the star's motion
with respect to other stars. Since the gas has a 10 \kms\ dispersion,
$5-10^{\circ}$ deviations can be expected, which would favour
$\vec{b}$. If the proposed path is correct, the origin of the binary
system would be on the outskirts of Vel~OB1.  The usually adopted
distance towards \vela\ of $1.9 \pm 0.1$~kpc (Sadakane et
al. \cite{Sa85}) is in accordance with the stellar parameters of
HD77581. The accretion induced X-ray luminosity of \vela\ (Kaper et
al. \cite{KH93}), also agrees very well with the distance towards
Vel~OB1. Thus, our observations confirm the proposed runaway status of
HD77581 and support its suggested origin in Vel~OB1.

In conclusion, the discovered wind bow shock around \vela\ provides
compelling observational evidence for a runaway high-mass X-ray
binary.  The direction of motion of the system is derived from the
symmetry of the bow shock and it is very likely that the OB
association Vel~OB1 is the origin of the system. From the observed
proper motion one can compute that the system left the outskirts of
Vel~OB1 about 2.5 million years ago, which corresponds to the expected
time of the supernova explosion of Vela~X-1's progenitor. Therefore,
our observations support Blaauw's scenario for the formation of the
runaway high-mass X-ray binary HD77581 (\vela ). It remains to be
shown whether all HMXBs are runaway systems. We predict that the
answer to this question is positive and expect that in at least a
third of the cases a wind bow shock is associated with them.

\acknowledgments

This paper is based on observations obtained at the 
European Southern Observatory in La Silla, Chile. We thank
J.~Bergeron, L.~Lucy and J.~van Paradijs for carefully reading the
manuscript. G.~Van de Steene produced the black-body fit to the
observed IRAS fluxes. The referee is acknowledged for his/her helpful remarks
and a critical reading of the manuscript.

\clearpage

\begin{table}
\caption[]{Proper motion, radial velocity, and space velocity of
HD77581 at the distance (1820 parsec) of Vel~OB1; for the space
velocity, the average radial velocity of the OB stars in
Vel~OB1 (26.7 \kms, Humphreys, \cite{Hu78}) is taken into account.
$\star$: radial velocity taken from Gies \& Bolton (\cite{GB86}).}
\begin{tabular}{l|llll} \hline\noalign{\smallskip}
 & $\mu_{\alpha} \cos{\delta}$ &  $\mu_{\delta}$ & $v_{r}$ & $v_{\star}$ \\
 & (''/yr) & (''/yr) & (\kms) & (\kms) \\ \hline \noalign{\smallskip}
(a) Turon et al. \cite{Tu92} & +0.004(3) & +0.009(3) & $-1$ & 89(40) \\ 
(b) Sahu \cite{Sa92}  & -0.0014(19) & +0.0107(22) & $-4^{\star}$ & 98(26) \\ 
\end{tabular}
\end{table}

\clearpage

{\it Fig. 1:} (Plate X): An R-band corrected \ha\ image of the field around
\vela\ is shown. North is up and east to the left. The images were
obtained with the DFOSC instrument and a CCD detector in the
Cassegrain focus of the 1.54m Danish telescope at the European Southern
Observatory. A wind bow shock is detected about 0.9 arcminute north of
the (saturated) 6$^{\mbox{th}}$ magnitude star HD77581. The western
(right) arm of the bow shock is split into two filaments, a feature
more often encountered in wind bow shocks.
\vspace{0.5cm}

{\it Fig. 2:} A flux-calibrated long-slit spectrum of the bow shock
was obtained with the Boller \& Chivens spectrograph mounted on the
1.52m ESO telescope. The spectrum was collapsed in the spatial
direction along two regions in the slit (length 4.2 arcminutes) which
are shown in the sub panel. The spectral resolution is 5.5~\AA\
(FWHM). Region A corresponds to the secondary filament which appears
to produce the maximum \ha\ intensity. Region B is the extension of
the wind bow shock to the west. The forbidden \mbox{[O~\sc{iii}]}
lines are indicative of shock excitation.
\vspace{0.5cm}

{\it Fig. 3:} The position of HD77581 (\vela) on the sky with respect
to the OB association Vel~OB1. Right ascension and declination are
given in degrees, north is up and east to the left. The direction of
proper motion, listed in the HIC ($\vec{a}$) and obtained by
Sahu (\cite{Sa92}) ($\vec{b}$), is shown. The proposed path of the system,
based on the symmetry axis of the bow shock, is given by a dotted
line. The numbers along this line indicate the position where the
system would have been (respectively 1,2, and 3 million years ago),
given a velocity of 90 \kms. 


\clearpage

\begin{thebibliography}{}
\bibitem[1971]{BK71}
Baranov, V.B., Krasnobaev, K.V., Kulikovskii, A.G. 1971, Sov.\
    Phys.\ Dokl.\ 15, 791
\bibitem[1961]{Bl61} 
Blaauw, A. 1961, \bain\ 15, 265
\bibitem[1993]{Bl93} 
Blaauw, A. 1993, in ASP Conf.\ Series, Volume 35, p. 207
\bibitem[1996]{DV96}
Dgani, R., Van Buren, D., Noriega-Crespo, A. 1996, ApJ 461, 372
\bibitem[1977]{Do77}
Dopita, M. 1977, \apjs\ 33, 437
\bibitem[1986]{GB86}
Gies, D.R., Bolton, C.T. 1986, \apjs\ 61, 419
\bibitem[1988]{HW88}
Helou, G., Walker, D.W. 1988, {\it IRAS Catalogs and Atlases}, NASA
    RP-1190, Vol.\ 7
\bibitem[1983]{Hi83}
Hills, J.G. 1983, \apj\ 267, 322
\bibitem[1992]{HO92}
Hollis, J.M., Oliversen, R.J., Wagner, R.M., Feibelman, W.A. 1992,
    \apj\ 93, 217
\bibitem[1978]{Hu78}
Humphreys, R.M. 1978, \apjs\ 38, 309
\bibitem[1973]{JL73}
Jones, C., Liller, W. 1973, \apjl\ 184, 121
\bibitem[1993]{KH93}
Kaper, L., Hammerschlag-Hensberge, G., Van Loon, J.Th. 1993, \aap\
    279, 485
\bibitem[1976]{MR76}
McClintock, J.E., Rappaport, S., Joss, P.C., et al. 1976, \apjl\ 206, 99
\bibitem[1989]{Na89}
Nagase, F. 1989, \pasj\ 41, 1
\bibitem[1989]{Os89}
Osterbrock, D.E. 1989, {\it Astrophysics of Gaseous Nebulae and AGN},
Univ.\ Sc.\ Books, CA
\bibitem[1996]{PE96}
Philp, C.J., Evans, C.R., Leonard, P.J.T., Frail, D.A. 1996, \aj\ 111, 1220
\bibitem[1967]{PR67}
Poveda, A., Ruiz, J., Allen, C. 1967, Bol.\ Obs.\ Tonantzintla y
    Tacubaya 4, 860
\bibitem[1985]{Sa85}
Sadakane, K., et al. 1985, \apj\ 288, 284
\bibitem[1992]{Sa92}
Sahu, M.S. 1992, Ph.D.\ Thesis University of Groningen
\bibitem[1992]{Tu92}
Turon, C., et al. 1992, The Hipparcos Input Catalogue, ESA SP-1136
\bibitem[1988]{BM88}
Van Buren, D., McCray, R. 1988, \apjl\ 329, 93
\bibitem[1995]{BN95} 
Van Buren, D., Noriega-Crespo, A., Dgani, R. 1995, \aj\ 110, 2914
\bibitem[1993]{He93}
Van den Heuvel, E.P.J. 1993, in {\it Saas-Fee Advanced Course on 
    Interacting Binaries} (Springer-Verlag), p. 263 
\bibitem[1995a]{KP95a}
Van Kerkwijk, M.H., Van Paradijs, J., Zuiderwijk, E.J., et al.
    1995a, \aap\ 303, 483
\bibitem[1995b]{KP95b}
Van Kerkwijk, M.H., Van Paradijs, J., Zuiderwijk, E.J. 1995b, \aap\
    303, 497
\bibitem[1989]{VO89}
Van Oijen, J.G.J. 1989, \aap\ 217, 115
\bibitem[1977]{PZ77}
Van Paradijs, J., Zuiderwijk, E.J., Takens, R.J., et al. 1977, \aap\
    30, 195
\bibitem[1996]{RV96}
Van Rensbergen, W., Vanbeveren, D., De Loore, C. 1996, \aap\ 305, 825
\bibitem[1973]{VW73}
Vidal, N.V., Wickramasinghe, D.T., Peterson, B.A. 1973, \apj\ 265, 1036
\bibitem[1977]{WG77}
Watson, M.G., Griffiths, R.E. 1977, \mnras\ 178, 513
\bibitem[1957]{Zw57}
Zwicky, F. 1957, {\it Morphological Astronomy} (Berlin: Springer), p. 258 
\end{thebibliography}
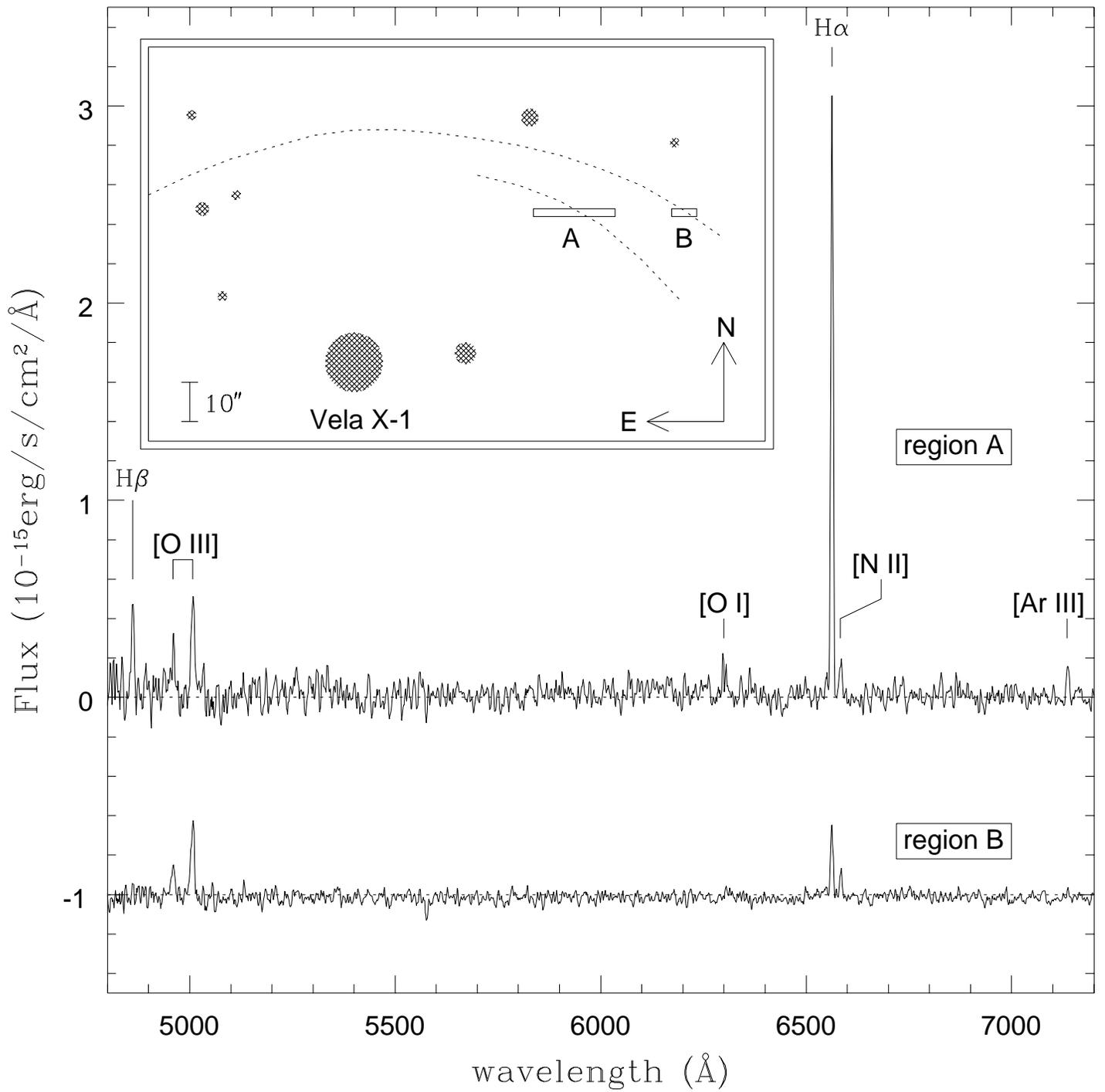
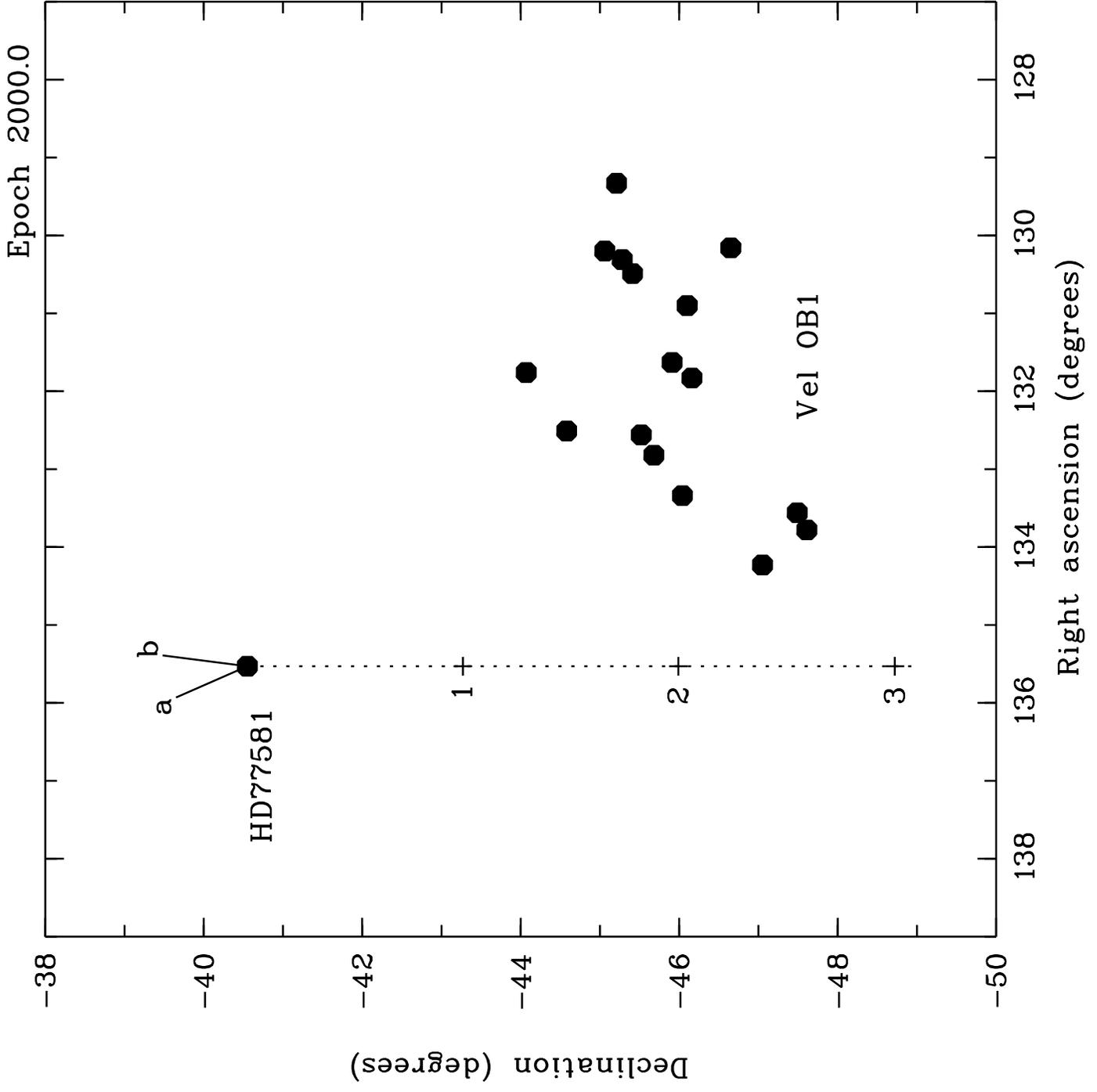
\end{document}